\documentclass[manuscript]{raa}           

\usepackage{graphicx,times}
\usepackage{natbib}
\usepackage{amssymb,amsmath}
\bibpunct{(}{)}{;}{a}{}{,}

\usepackage[pagebackref=true]{hyperref}

\begin{document}

   \title{The Potential of Detecting Radio-flaring Ultracool Dwarfs at $L$ band in the FAST Drift-scan Survey}

 \volnopage{ {\bf 20XX} Vol.\ {\bf X} No. {\bf XX}, 000--000}
   \setcounter{page}{1}

   \author{Jing Tang
   \inst{1}, Chao-Wei Tsai\inst{1}, Di Li\inst{1,2,3}
   }

   \institute{ CAS Key Laboratory of FAST, National Astronomical Observatories, Chinese Academy of Sciences, Beijing 100101, China; {\it jtang@nao.cas.cn, cwtsai@nao.cas.cn, dili@nao.cas.cn}\\
        \and
             University of Chinese Academy of Sciences, Beijing 100049, China\\
	\and
NAOC-UKZN Computational Astrophysics Centre, University of KwaZulu-Natal, Durban 4000, South Africa\\
\vs \no
   {\small Received 2022 January 7; revised 2022 April 1; accepted 2022 April 5}
}

\abstract{The Five-hundred-meter Aperture Spherical radio Telescope (FAST) completed its commissioning and began the Commensal Radio Astronomy FasT Survey (CRAFTS), a multi-year survey to cover 60\% of the sky, in 2020. We present predictions for the number of radio-flaring ultracool dwarfs (UCDs) that are likely to be detected by CRAFTS. Based on the observed flaring UCDs from a number of unbiased, targeted radio surveys in the literature, we derive a  detection rate of $\ge$3\%. Assuming a flat radio spectrum  $\nu L _{\nu}\propto \nu^{\beta+1} $ with $\beta$ = -1.0 for UCD flares, we construct a flare luminosity function $d N/d L \propto L^{-1.96 \pm 0.45}$ (here $L=\nu L_\nu$).
CRAFTS is found to be sensitive enough for flares from UCDs up to $\sim$180 pc. Considering the Galactic thin disk, we carry out a 3D Monte Carlo simulation of the UCD population, which is then fed to mock CRAFTS observations. We estimate that $\sim$170 flaring UCDs would be detected through transient searches in circular polarization. Though only marginally sensitive to the scale height of UCDs, the results are very sensitive to the assumed spectral index $\beta$. For $\beta$ from 0 to -2.5, the number of expected detections increases dramatically from $\sim$20 to $\sim$3460. We also contemplate the strategies for following up candidates of flaring UCDs, and discuss the implications of survey results for improving our knowledge of UCD behavior at $L$ band and dynamos. 
}

   \authorrunning{J. Tang et al. }            
   \titlerunning{Radio-flaring UCDs}  
   \maketitle

%
\section{Introduction}           
\label{sec:intro}
Ultracool dwarfs (UCDs) include very low mass stars (spectral types $\ge$ M7) and brown dwarfs (L, T and Y) \citep{1999ApJ...519..802K}. Since the discovery of the unexpected radio emission from LP 944--20 (M9) in 2001 \citep{2001Natur.410..338B}, more radio-active UCDs have been found.
This radio emission is non-thermal and can be used as the magnetic activity indicator. For solar-type stars, magnetic fields are thought to depend on the shearing at the interface between the inner radiative zone and the outer convective envelope \citep{1975ApJ...198..205P}. In contrast, UCDs are fully convective. How magnetic fields are generated and maintained in these objects is a big challenge to our understanding of dynamo processes. There are already many dynamo models (e.g. \citealt{2008ApJ...676.1262B, 2009Natur.457..167C, 2015ApJ...813L..31Y}), and more observational evidences are needed to test these models. For earlier type stars, H$\alpha$ emission from the chromosphere and X-ray emission from the corona are also used to indirectly trace the magnetic activity. However, these emissions decline significantly in UCDs \citep{2000AJ....120.1085G, 2003ApJ...583..451M, 2004AJ....128..426W, 2010ApJ...709..332B, 2014ApJ...785....9W, 2015AJ....149..158S}, and thus are much less efficient probes in these objects. 

Radio-active UCDs exhibit both flaring and quiescent (also known as persistent) radio emission. The flaring emission is thought to be caused by the electron cyclotron maser (ECM) instability (\citealt{2008ApJ...684..644H}; see \citealt{2006A&ARv..13..229T} for a review). \citet{2015Natur.523..568H} further detected both radio and optical emissions powered by magnetospheric currents from LSR J1835+3259 (M8.5), confirming this coherent radio emission has the auroral origin as that from magnetized planets in solar system. Apart from the auroral origin which requires large-scale fields, this ECM flaring emission can be also produced in localized active regions \citep{2006A&ARv..13..229T, 2015ApJ...802..106L}, similar to solar and stellar flares attributed to maser action. The ECM emission is highly circularly polarized (up to 100\%) \citep{2005ApJ...626..486B, 2007ApJ...663L..25H, 2008ApJ...684..644H, 2008AA...487..317A, 2009ApJ...695..310B, 2012ApJ...747L..22R, 2015ApJ...799..192W, 2015ApJ...808..189W, 2016ApJ...821L..21R} and rotationally modulated \citep{2006ApJ...653..690H, 2007ApJ...663L..25H, 2008ApJ...684..644H, 2009ApJ...695..310B, 2015ApJ...808..189W, 2018ApJS..237...25K}. This emission is produced at the electron cyclotron fundamental frequency \citep{2006A&ARv..13..229T}, $\nu_{\rm MHz} \approx 2.8 \times B_{\rm Gauss}$, and can be used to measure local magnetic field strengths. As the magnetic field on a UCD extends into a range of heights over the surface, the bandwidth is then determined by the difference between the highest and lowest field strengths. Multi-frequency observations are needed to study the nature of the radio emission. The quiescent emission is considered to be gyrosynchrotron or synchrotron radiation from mildly or highly relativistic electrons \citep{2006ApJ...653..690H, 2015ApJ...815...64W, 2017ApJ...846...75P}.

Among the dynamo models for fully convective stars, \citet{2009Natur.457..167C} claimed convected energy flux determines the magnetic field strength and proposed a convection-driven dynamo scaling law for describing field strengths of rapidly rotating low-mass stars and planets. Based on that, \citet{2009ApJ...697..373R} provided a semi-empirical method
to estimate the field strengths of brown dwarfs and giant planets from their masses, radii, and luminosities. However, \citet{2018ApJS..237...25K} showed a clear departure of T dwarfs from this scaling law. They gave several possible explanations, such as higher-order non-dipole fields and age-evolving effect on magnetic energy. Yet the conclusions cannot be drawn with the small number of observed cases. More samples of late L and T dwarfs are needed to investigate the field topologies and the age dependence in dynamos. On the other hand, \citet{2019MNRAS.487.1994K} took pilot observations of three Y dwarfs with VLA at 4-8 GHz and failed to detect any pulsed radio emission. One possibility is that the observed frequency is too high, as the predicted field strengths for Y dwarfs are $\sim$hundreds of gauss \citep{2009Natur.457..167C}, and the pulse caused by ECM should be lower than 4 GHz. Thus surveys at lower frequencies are crucial to search for new objects with weaker magnetic fields and shed light on convective dynamos.

The VLA Faint Images of the Radio Sky at Twenty-cm (FIRST; \citealt{1995ApJ...450..559B}) survey is a well-known 1.4 GHz survey with high sensitivity ($\sim$0.15 mJy) and resolution ($\sim$5$"$). This survey has a point-source detection threshold $\sim$1 mJy, while covering $\sim$10,000 deg$^2$. A comprehensive search for radio-active UCDs in the FIRST survey is not conducted. Although most known radio-active UCDs are outside the FIRST coverage, \citet{2011ApJ...741...27M} identified the persistent, non-flaring emission from 2MASS J13142039+1320011 in this survey.
Another noticeable 1.4 GHz survey, the NRAO VLA Sky Survey (NVSS; \citealt{1998AJ....115.1693C}) covers the sky north of $\delta$ = -40$^\circ$, but the detection limit is $\sim$2.5 mJy, not sensitive enough to detect the predicted 1.4 GHz radio emission from UCDs. Recently, in the LOFAR Two-meter Sky Survey (LoTSS; \citealt{2017A&A...598A.104S, 2019A&A...622A...1S}), \citet{2020ApJ...903L..33V} discovered a T6.5 dwarf BDR J1750+3809, indicating that low-frequency radio surveys can discover substellar objects that are too cold and/or distant to be detected in current infrared surveys.

FAST, the largest single dish telescope in the world, will be able to efficiently investigate the radio-flaring UCDs at low frequencies ($\le$1.4 GHz). The telescope finished its commissioning in early 2020. FAST has begun the Commensal Radio Astronomy FasT Survey (CRAFTS; \citealt{2018IMMag..19..112L}),
which utilizes a novel and unprecedented mode to realize simultaneous data taking for pulsar search, Galactic H\,{\sc i} mapping, H\,{\sc i} galaxy study and transient FRB search. CRAFTS implements a high frequency CAL injection scheme, which facilitates the recovery of bandpass and the monitoring of polarization calibration. 

We consider here the advantages and challenges of radio-flaring UCD detections through the CRAFTS survey. For FAST, the most significant problem is the severe confusion due to the large beam size. The highly circularly polarized nature of the radio flaring emission, in conjunction with a full polarization recording should improve significantly their detectability. Also, CRAFTS is designed to be a two-pass survey, the flux comparison between two epochs can be used to search for possible radio transients. Increasing radio-active UCD sample would advance our understanding of physical properties (e.g. age, rotation rate and bolometric luminosity) associated with magnetic activities in these objects.  

In this paper, we provide the predictions on the potential of FAST to detect radio flares from UCDs. In Section~\ref{parameters}, we investigate the flaring UCD detection rate, the radio luminosity function (LF) and the duty cycle of the flares, which are the necessary ingredients for the following simulation. In Section~\ref{potential}, we describe CRAFTS specifications and carry out the Monte Carlo simulation to explore the possibility of searching for flaring UCDs in CRAFTS. In Section~\ref{discuss}, we consider the variation of each ingredient, including the spectral index of the flaring emission, in the simulation and their impact on the results. We also discuss the target selection strategy for follow-up observations, and the implication of flaring UCD search in CRAFTS. Finally we present our conclusions in Section~\ref{conclusion}.


\section{Relevant parameters} \label{parameters}

In this section, we summarize the radio-active UCDs detected in previous surveys from the literature, and assess the flaring UCD detection rate, construct the LF of the radio flares, and estimate the flare duty cycle. These parameters are kernels of the Monte Carlo simulation in Section~\ref{mcs} to predict the number of radio-flaring UCDs that could potentially be discovered in CRAFTS. Our simulation follows the methodology of \citet{2017ApJ...845...66R}, yet includes more careful treatment to derive the radio LF of UCD flares based on the results of previous radio UCD surveys, though the sample is very small. The duty cycle and the temporal evolution of the flare which were overlooked in \citet{2017ApJ...845...66R} are also included.

\subsection{Previous surveys for radio-active UCDs} \label{surveys}

Over the past two decades, a number of surveys have been carried out to search for radio-active UCDs, motivated by the first radio detection of LP 944--20 \citep{2001Natur.410..338B}. Till today, there are 26 detections of radio-emitting UCDs reported in the literature. The properties of these sources were summarized in \citet{2017ApJ...846...75P}, except VHS 1256-1257 \citep{2018A&A...610A..23G} and 2MASS J17502484-0016151 \citep{2020ApJ...903...74R} . 

\citet{2002ApJ...572..503B} utilized VLA to observe 11 dwarfs ranging from M8 through T6 and detected flares and persistent emission from three sources. \citet{2005ApJ...626..486B} monitored seven Southern late M and L dwarfs with ATCA at 4.80 and 8.64 GHz simultaneously, and two sources were found to exhibit radio emission, one of which is the most radio-luminous UCD to date, namely DENIS J1048-3956. \citet{2006ApJ...648..629B} analyzed a large sample of 88 objects ranging from M7 to T8 observed by VLA, and found two sources showed quiescent emission. \citet{2006ApJ...644L..67O} observed three young UCDs (M8--M8.5) with no radio detection. Later \citet{2012ApJ...746...23M} conducted another survey with VLA, including 76 dwarfs of spectral types M7--L3.5, and detected radio emission from two UCDs. In addition, \citet{2007ApJ...658..553P} took VLA observations of eight UCDs in a narrow range of spectral type M8--M9.5 and found one more quiescent source. The aforementioned surveys were all conducted at 8.5 GHz. In order to search for UCDs capable of producing radio emission at lower frequencies, \citet{2008AA...487..317A} attempted a mini-survey of eight dwarfs spanning from M8.5 to T6 with VLA at 4.9 GHz, contributing one new flaring source, while a later survey of 32 UCDs covering spectral type M7--T8 made no detection \citep{2013A&A...549A.131A}. Moreover, \citet{2014ApJ...785....9W} observed five UCDs (M7.5--M9) with upgraded VLA centered at 5.0 and 7.1 GHz simultaneously, yielding one detection. Later \citet{2016MNRAS.457.1224L} did a small ATCA survey of Southern 15 UCDs ranging from M7 to L8 centered at 5.5 and 9.0 GHz simultaneously, and found a new quiescent source. More recently, \citet{2019MNRAS.483..614Z} analyzed GMRT observations of nine UCDs (mainly L dwarfs) at around 1300 and 610 MHz, without any new detection. On the other hand, a pilot survey with VLA C-band targeting H$\alpha$-emitting  and/or optical/IR variable dwarfs (L7--T6.5), achieved detection in four of five objects \citep{2016ApJ...818...24K}. This target selection strategy is based on the work of \citet{2015Natur.523..568H} who suggested the highly circularly polarized radio emission has the same nature as the auroral emission from magnetic planets in solar system and is driven by large-scale currents due to the magnetosphere-ionosphere coupling. Furthermore, \citet{2019MNRAS.487.1994K} observed three known IR-variable Y dwarfs, but without any new detection. \citet{2020ApJ...903...74R} made a survey of 17 photometrically variable brown dwarfs with VLA at 4-8 GHz, and found one more source with both quiescent and flaring emissions. In addition, there were several other radio-detected UCDs in the targeted observations \citep{2013ApJ...762L...3B, 2013ApJ...779..172G, 2015AJ....150..180B, 2015ApJ...805L...3O, 2016AJ....152..123G, 2018A&A...610A..23G}.

Apart from interferometers, Arecibo was also used to study radio-flaring UCDs. \citet{2013ApJ...773...18R} initially led the 4.75 GHz Arecibo survey of 33 objects spanning spectral range M9--T9, and discovered the first radio-emitting T dwarf. Another flaring T dwarf was detected in the second survey of 27 UCDs encompassing spectral type M7--T9 \citep{2016ApJ...830...85R}. 

In total, 26 out of 269 observed UCDs, including a few that were surveyed more than once, were detected as radio emitters, yielding an overall detection efficiency of $\sim$10\%. It is worth noting that the biased survey in \citet{2016ApJ...818...24K} guided by auroral activity tracers such as H$\alpha$ emission and/or optical/IR variability seems to be efficient. However, other optical H$\alpha$ surveys may also result in low detection rate. For example, \citet{2016ApJ...826...73P} conducted an optical spectroscopic survey of  L and T dwarfs with the Keck telescopes and achieved an H$\alpha$ detection rate of $\sim$9.2\%, statistically equivalent to that of the radio surveys aforementioned. Another thing we would like to point out is that BDR J1750+3809 discovered in LoTSS \citep{2020ApJ...903L..33V} is the only radio-active brown dwarf reported in that blind survey so far, and the detection rate is unknown. Hence we do not consider this object in the following statistical analysis.

\subsection{Detection rate of radio-flaring UCDs from unbiased surveys} \label{rate}

As FAST is sensitive to rapidly varying radio emission, here we have only estimated the detection rate of flaring UCDs from those unbiased, targeted surveys since 2002 
\citep{2002ApJ...572..503B, 2006ApJ...648..629B, 2005ApJ...626..486B, 2007ApJ...658..553P, 2008AA...487..317A, 2013A&A...549A.131A, 2012ApJ...746...23M, 2013ApJ...773...18R, 2016ApJ...830...85R, 2014ApJ...785....9W, 2016MNRAS.457.1224L, 2019MNRAS.483..614Z}. Here the term `unbiased' means that the targets are selected without preferences, e.g. rotation, age, or X-ray and H$\alpha$ activity, which are thought to be associated with the radio activity. 
These surveys were carried out using different observation modes, however, because they are sensitive to several-minute-long flares, we still combine these results for statistical analysis. In these surveys, a total of 236 UCDs of spectral types M7-T9 were observed, and seven radio-flaring dwarfs were detected (see Table~\ref{tab1}), yielding a detection rate of $\sim$3\%. 

Flaring UCDs cannot be always identified in the surveys, e.g. LSR J1835+ 3259 and 2MASS J1314+1320. The quiescent emission was initially detected from these objects in \citet{2006ApJ...648..629B} and \citet{2012ApJ...746...23M}, and then follow-up observations confirmed the occurrence of flares \citep{2008ApJ...684..644H, 2015ApJ...799..192W}. This may be due to insufficient observing time. Another possibility is the change of radio emission properties, which are related to magnetic structures, e.g. solar-like activity cycles originated from magnetic reversals \citep{2016ApJ...830L..27R} may cause flaring UCDs to undergo low magnetic activity periods. The low detection rate can be also caused by other observational selection effects. For example, only when the beamed radio emission sweeps in the direction of the Earth during one rotation period of an object, it can be detected. Thus those with long rotational periods have much lower chances to be detected. Additionally, previous work focused on radio observations at $\sim$5 GHz and $\sim$8 GHz. That means objects with field strengths lower than $\sim$1.5 kGauss cannot be detected if ECM mechanism operates. On the other hand, this number may imply that radio inactive dwarfs occupy a large proportion, owing to the absence of either the magnetic field or the electrodynamic engine (e.g. satellite-induced, similar to Jupitor-Io system) to generate the radio emission \citep{2015Natur.523..568H, 2016ApJ...818...24K, 2017ApJ...846...75P}.

It is worth noting that in the radio surveys of UCDs with interferometers, \citet{2006ApJ...648..629B}, \citet{2007ApJ...658..553P}, \citet{2013A&A...549A.131A}, \citet{2014ApJ...785....9W} and  \citet{2016MNRAS.457.1224L} searched for variability (flares) just for the objects that were detected in the images, and \citet{2012ApJ...746...23M} only examined the images for all sources. This could potentially introduce an issue of lower detection rate due to the flux dilution. When a flare with duration $\Delta t$ occurs in the observation of time $T$ ($\Delta t < T$), the signal-to-noise ratio (SNR) for the flare would be higher than that for the entire observation. We take the observation of TVLM 513-46 in \citet{2007ApJ...663L..25H} as an example. Referring to the phase-folded light curve at 8.44 GHz \citep[see Figure~3 in][]{2007ApJ...663L..25H}, a single burst lasts for $\sim$8 min with the peak flux of $\sim$4 mJy, and the mean flux over a period of 1.958 hr is 0.464 mJy. The rms noise grows with $time^{1/2}$, while the signal grows with $time$. Thus the SNR of the burst ($\propto 4 \times 8/\sqrt{8}$) exceeds the SNR of the duration of one rotation period ($\propto 0.464 \times 1.958 \times 60/\sqrt{1.958 \times 60}$) by a factor of $\sim$2.2. Here we use the peak flux instead of the average flux of the burst over the $\sim$8 min, and the resultant value would be slightly larger. \citet{2008AA...487..317A} reexamined the $X$ band data in \citet{2007ApJ...663L..25H} and obtained the value of 1.5. Note that if the quiescent emission is very low or absent, then the SNR of the burst over the whole observation would be diluted by a factor of $(\Delta t/T)^{1/2}$. Further, for the radio detection determined from the image, light curve needs to be constructed to search for the variability to confirm the fidelity of the flaring event
\citep[e.g.][for J0952219 -- 192431]{2012ApJ...746...23M}. Thus, if only 
synthesized interferometric images were made to search for radio emission without time series analysis, flares may be missed. With this concern in mind, we consider $\sim$3\% as a lower limit of the detection rate. As there is no obvious evidence that the detection rate changes across spectral types \citep{2016ApJ...830...85R}, we use a single rate 3\% for all types of UCDs in the simulation. 

\begin{table}
\begin{minipage}[]{100mm}
\caption[]{UCDs with Radio Flares from unbiased surveys\label{tab1}}
\end{minipage}
\setlength{\tabcolsep}{2.5pt}
\small
\centering
 \begin{tabular}{llcccccccccl}
  \hline\noalign{\smallskip}
Object& SpT& $d$& [$L_{\rm bol}$]& $\nu_{\rm obs}$& $t_{\rm on-source}$& $\tau_{\rm flare}$& $F_{\rm \nu, R}$& $\nu L_{\nu}$& Period& Size& References$^{\rm b}$\\
& &(pc)& $L_{\odot}$)& (GHz)& (ks)& (min)& (mJy)& (10$^{24}$ erg s$^{-1}$)& (hr)$^{\rm a}$& & \\
  \hline\noalign{\smallskip}
J1048-3956$^{\rm c}$&M8.5&4.05 $\pm$ 0.002&-3.51&4.80&36&4--5$^{\rm d}$&6.00 $\pm$ 0.90&0.57 $\pm$ 0.08&...&7&14, 10, 9, \bf{6}\\
&&&&8.64&36&4--5$^{\rm d}$&29.00 $\pm$ 1.00&4.92 $\pm$ 0.17&&\\
TVLM 513-46&M8.5&10.70 $\pm$ 0.02&-3.57&8.46&6.6&13.8$^{\rm e}$&0.89 $\pm$ 0.09&1.03 $\pm$ 0.10&1.96&12&15, 10, 9, \bf{2}, \underline{11}\\ 
J0024-0158$^{\rm c}$&M9.5&12.51 $\pm$ 0.03&-3.44&8.46&5.5&5.6$^{\rm e}$&0.33 $\pm$ 0.09&0.52 $\pm$ 0.14&...&12&18, 10, 9, \bf{2}\\
J0746+2000 B$^{\rm c, f}$&L1.5&12.36 $\pm$ 0.12&-3.77&4.90&5.4$^{\rm g}$&3.4$^{\rm h}$&1.60 $\pm$ 0.30$^{\rm h}$&1.43 $\pm$ 0.27&2.07&8&5, 8, 13, \bf{1}, \underline{3}\\
J0036+1821$^{\rm c}$&L3.5&8.74 $\pm$ 0.02&-3.93&8.46&9.6&21.5$^{\rm e}$&0.54 $\pm$ 0.10&0.42 $\pm$ 0.08&3.08&12&16, 10, 9, \bf{2}, \underline{12}\\
J1122+2550$^{\rm c}$&T6&17.72 $\pm$ 1.59&-4.90&4.75&7.2&2.0$^{\rm i}$&1.50 $\pm$ 0.20$^{\rm j, k}$&2.68 $\pm$ 0.60&...$^{\rm l}$&20$^{\rm m}$&17, 4, 19, \bf{21, 22}\\
J1047+2124$^{\rm c}$&T6.5&10.56 $\pm$ 0.42&-5.30&4.75&3.6&0.8&2.70 $\pm$ 0.20$^{\rm j}$&1.71 $\pm$ 0.19&1.77&31$^{\rm n}$&7, 23, 9, \bf{20}, \underline{24}\\
  \noalign{\smallskip}\hline
\end{tabular}
\tablecomments{1.0\textwidth}{The columns are (left to right) (1) object name; (2) spectral type; (3) distance; (4) bolometric luminosity; (5) observing frequency; (6) total on-source observing time; (7) flare duration; (8) peak flux density; (9) radio luminosity; (10) period; (11) the size of the observed UCD sample in the survey; (12) references. \\
$^{\rm a}$ The periods are obtained from the radio data.\\
$^{\rm b}$ The references in sequence denote the source for the spectral type, the distance, and the bolometric luminosity. The radio survey and period references are in bold and underlined, respectively.\\
$^{\rm c}$ The full names of these objects are, in table order, DENIS J1048.0–3956, 2MASSI J0024246–015819, 2MASSI J0746425+200032, 2MASS J00361617+1821104, WISEP J112254.73+255021.5, and 2MASS J10475385 +2124234.\\ 
$^{\rm d}$ These two flares were detected at 4.8 GHz and 8.46 GHz simultaneously, with $\sim$10 min apart.\\
$^{\rm e}$ These values are flare FWHM duration. For other sources in the table, the flare timescale is determined when the envelope of the flare departs from the background emission.  \\  
$^{\rm f}$ J0746+2000 is an L0+L1.5 binary system \citep{2004AA...423..341B}. The detected period of the radio pulses is consistent with that of the secondary L1.5 dwarf \citep{2009ApJ...695..310B, 2020ApJ...897...11Z}. Note that the primary is also radio emitter \citep{2020ApJ...897...11Z}.\\ 
$^{\rm g}$ This value is calculated according to the archive data at \url{https://archive.nrao.edu/archive/advquery.jsp}. \\ 
$^{\rm h}$ The flux and duration correspond to the burst at 4.6 h in the light curve with 10 s time resolution \citep{2008AA...487..317A}. The duration is calculated when the envelope of the burst departs from the value of the quiescent component, $\sim$0.2 mJy. \\ 
$^{\rm i}$ The duration of the pulse is not clearly given in \citet{2016ApJ...821L..21R}, but it mentioned the detected pulses were 30-120 s. We choose 120 s as the duration of the first discovered flare according to its light curve.\\
$^{\rm j}$ There are multiple pulses recorded in the follow-up observations after the initial discovery of this flaring brown dwarf. For the purpose of selecting the unbiased flare measurement as the representation of the radio luminosity of this source, we use the flux detected in the original discovery.\\
$^{\rm k}$ The uncertainty is not clearly given in \citet{2016ApJ...821L..21R}. As the observations were carried out in the same way as \citet{2013ApJ...773...18R}, we adopt the same uncertainty.\\
$^{\rm l}$ \citet{2016ApJ...821L..21R} proposed a period of $\sim$17.3 min, while \citet{2017ApJ...834..117W} inferred the period of $\sim$116 min.\\
$^{\rm m}$ In the survey of 27 brown dwarfs, 20 are not-previously observed targets by Arecibo.\\
$^{\rm n}$ In the survey of 33 brown dwarfs, TVLM 513-46 and J0746+2000 are included as calibration sources.\\
References. (1) \citet{2008AA...487..317A}, (2) \citet{2002ApJ...572..503B}, (3) \citet{2009ApJ...695..310B}, (4) \citet{2018ApJS..234....1B}, (5) \citet{2004AA...423..341B}, (6) \citet{2005ApJ...626..486B}, (7) \citet{2006ApJ...637.1067B}, (8) \citet{2017ApJS..231...15D}, (9) \citet{2015ApJ...810..158F}, (10) \citet{2018AA...616A...1G}, (11) \citet{2007ApJ...663L..25H}, (12) \citet{2008ApJ...684..644H}, (13) \citet{2013AA...554A.113H}, (14) \citet{2004AJ....128.2460H}, (15) \citet{1995AJ....109..797K}, (16) \citet{2000AJ....120..447K}, (17) \citet{2011ApJS..197...19K}, (18) \citet{2001ApJ...548..908L}, (19) \citet{2017ApJ...846...75P}, (20) \citet{2013PhDT.......226R}, (21) \citet{2016ApJ...821L..21R}, (22) \citet{2016ApJ...830...85R}, (23) \citet{2004AJ....127.2948V}, (24) \citet{2015ApJ...808..189W}.
}
\end{table}

\subsection{Radio luminosity function of UCD flares} \label{lf}

In this section, we build a sample to derive the $L$-band LF of UCD flares.
We combine the results from previous small size surveys, and establish a uniform standard to recalculate the flux densities of the flaring events from these inhomogeneous surveys. 

For the radio flares of UCDs -- J1048-3956, TVLM 513-46, J0024-0158, J0746+2000 and J0036+1821 -- that were first detected by interferometers, the peak flux densities are adopted directly from the light curves in the literature \citep{2005ApJ...626..486B, 2002ApJ...572..503B, 2008AA...487..317A}. The median of the data outside the flare periods is calculated as the quiescent component, and the median absolute deviation of these data is then converted to the standard deviation to be considered as the flux uncertainty. The difference between the peak flux and the quiescent component is treated as the pure flare flux, and is used to construct the flare LF. We note that, for J1048-3956, the quiescent component, shown in Figure 3 of
\citet{2005ApJ...626..486B}, is estimated as 0.53 mJy at 4.80 GHz and 0.64 mJy at 8.64 GHz, while it is 0.14 mJy at 4.80 GHz and $<$ 0.11 mJy at 8.64 GHz reported in the literature. This difference is likely due to that the observation is 12 hr long, while the light curves provided in the literature were only for half an hour. For the case of J0746+2000, a series of bursts at 4.3 hr (two peaks), 4.5 hr, 4.6 hr and 4.9 hr are seen in the light curve, which is shown in top-left panel of Figure 2 in \citet{2008AA...487..317A}. Here the median of these burst fluxes as the peak is adopted. For the detection of TVLM 513-46, J0024-0158 and J0036+1821 in \citet{2002ApJ...572..503B}, the time resolution in their light curves is 5.5 min, 8 min and 6.5 min, respectively. These large time bins may exceed the timescale of a single flare. For sources J1122+2550 and J1047+2124, they were first detected by Arecibo, and the quiescent emission could not be distinguished
from those observations. The follow-up VLA observations, however, showed the quiescent emission of these two objects is two orders of magnitude fainter than their flares \citep{2013ApJ...767L..30W, 2015ApJ...808..189W, 2017ApJ...834..117W}, and thus the weak quiescent flux is neglected. The summary of these results are presented in Table~\ref{tab1}.

These flares were detected at $\sim$5 GHz or $\sim$8 GHz. In this work we assume the flux density spectra is $S_{\nu} \propto \nu^{\beta}$ with $\beta$ = -1.0. This is equivalent to the assumption of a flat $\nu L_{\nu}$ for the flaring emission at different frequencies. Detailed discussion on the spectral index of the flaring emission can be found in Section \ref{spectralindex}. We then assume the LF follows a power-law distribution,
\begin{equation}
dN = kL^{-\alpha}dL,
\end{equation}
where $dN$ is the number of flares with radio luminosity between $L$ and $L+dL$ (here $L = \nu L_{\nu}$), $k$ is the normalization constant, and $\alpha$ is the power-law index.

In fact, $dN/dL = kL^{-\alpha}$ is the probability density function of $L$, denoted by $p(L)$. We follow the recipe laid out in \citet{doi:10.1137/070710111} and estimate $\alpha$ using the method of \textit{maximum likelihood}. We write down the log-likelihood function,
\begin{equation}
{\rm ln} \mathcal{L} (\alpha) = n {\rm ln} k - \alpha \sum_{i=1}^{n}{\rm ln} L_i,
\label{mle}
\end{equation}
with the normalization constant $k = (\alpha-1) L_{\rm min}^{\alpha-1}$ for $\alpha >$ 1. Here $L_{\rm min}$ is the lower bound on the peak luminosities of radio flares. As mentioned in Section~\ref{rate}, 236 UCDs were observed in unbiased, targeted radio surveys, and seven flaring objects were detected with the minimal radio luminosity of 4.2 $\times$ 10$^{23}$ erg s$^{-1}$. In the remaining objects, about three-quarters have the detection limit below this luminosity. For a given LF with a positive $\alpha$, the chance of detecting flares with lower luminosities should be higher, yet there is no detection. Thus here we set $L_{\rm min}$ = 4.2 $\times$ 10$^{23}$ erg s$^{-1}$. By maximizing Equation~\ref{mle}, $\alpha$ can be derived \citep{doi:10.1137/070710111}, 
\begin{equation}
\alpha = 1 + n \Big[\sum_{i=1}^{n} {\rm ln}\frac{L_i}{L_{\rm min}} \Big]^{-1}.
\end{equation}
With our sample of flaring events reported in Table~\ref{tab1}, we obtain $\alpha$ = 1.96.

\begin{figure}
 \includegraphics[width=\columnwidth]{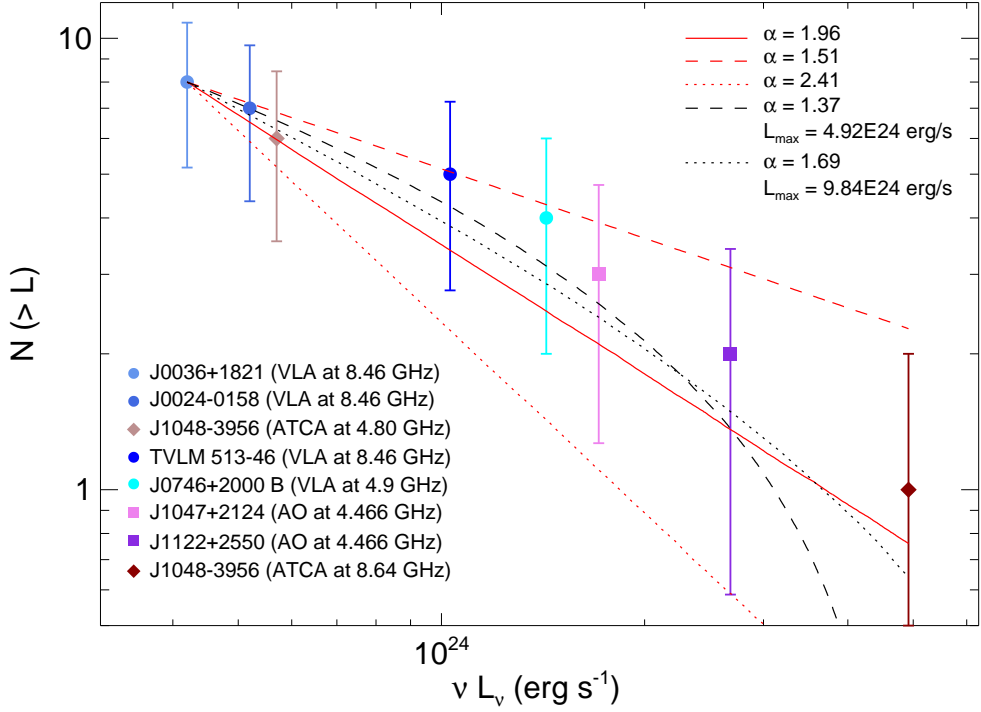}
 \caption{The power-law fits to the cumulative distribution of flaring events listed in Table~\ref{tab1}. The expected uncertainty of the cumulative distribution is $\sigma_{\rm{cum},i} = \sqrt{N_i}$ \citep{2015ApJ...814...19A}. The power laws are $\alpha$ = 1.96 (red solid line), $\alpha$ = 1.51 (red dashed line), $\alpha$ = 2.41 (red dotted line), $\alpha$ = 1.37 with $L_{\rm max}$ = 4.92 $\times$ 10$^{24}$ erg s$^{-1}$ (black dashed line) and $\alpha$ = 1.69 with $L_{\rm max}$ = 9.84 $\times$ 10$^{24}$ erg s$^{-1}$ (black dotted line). The latter two cases with high luminosity cutoff are discussed in Section~\ref{sec:parameter}. The lower bound $L_{\rm min}$ = 4.2 $\times$ 10$^{23}$ erg s$^{-1}$ is set for all power laws. \label{lf_plot}}
\end{figure}

It is worth noting that the derived LF suffers from the small sample size significantly. However, this is the best sample available from all published literatures so far. \citet{doi:10.1137/070710111} discussed the uncertainty of $\alpha$ derived from small sample in Appendix B, that is, $\sigma_{\alpha} = n(\alpha - 1)/(n-1)\sqrt{n-2}$. When $n$ is large, $\sigma_{\alpha} = (\alpha - 1)/\sqrt{n}$. In our case, $\sigma_{\alpha}$ = 0.45. The LF can be determined more accurately when more samples are available. The power-law fits with different parameters to the data are shown in Figure~\ref{lf_plot}. We also consider the impact of measurement uncertainties of $L$ on $\alpha$. Then the probability density function of $L$ is $p(L) = k/(\sqrt{2\pi}\sigma_L)\int_{L_{\rm min}}^\infty x^{-\alpha} {\rm exp}\big(-(x-L)^2/2\sigma_L^2\big)dx$. To maximize the log-likelihood function, we obtain the numerical solution $\alpha$ = 1.98. Thus this impact is negligible. In our simulation, we use the LF $dN/dL \propto L^{-1.96}$ with $L_{\rm min}$ = 4.2 $\times$ 10$^{23}$ erg s$^{-1}$. The uncertainty of $\alpha$ is also considered in the simulation (Section~\ref{mcs}). 

\subsection{Rotation periods, flare duty cycle and temporal evolution}\label{sec:duty_cycle}

The flaring radio emission is shown to be rotationally modulated \citep{2006ApJ...653..690H, 2007ApJ...663L..25H, 2008ApJ...684..644H, 2009ApJ...695..310B, 2015ApJ...808..189W, 2018ApJS..237...25K}. It is thus necessary to specify the distribution of rotation periods for the modeled UCD population. \citet{2014ApJ...793...75R} assumed a lognormal period distribution in the form of 
\begin{equation}
f(P)=\frac{1}{P \sigma_0 \sqrt{2 \pi}} e^{-\frac{({\rm ln} P-\mu_0)^2}{2 \sigma_0^2}},
\label{period}
\end{equation}
where $P$ is the period in hours, and the parameters $\mu_0$ = 1.41 and $\sigma_0$ = 0.48 are determined based on the previous $v~{\rm sin}~i$ studies of L and T dwarfs \citep{2006ApJ...647.1405Z, 2008ApJ...684.1390R}. \citet{2017ApJ...845...66R} reckoned the objects with rotation periods of $\le$1 hr and $\ge$ 10 hr may be underestimated \citep{2016ApJ...821L..21R, 2015ApJ...799..154M}, and suggested $\mu_0$ = 1.22 and $\sigma_0$ = 0.49. However, for WISEPC J112254.73 +255021.5, in contrast to the period of 17.3 min given in \citet{2016ApJ...821L..21R}, \citet{2017ApJ...834..117W} inferred a longer period of 116 min. Since there is no further discussion on the parameter issue as claimed in \citet{2017ApJ...845...66R}, here we adopt $\mu_0$ = 1.41 and $\sigma_0$ = 0.48.

The duty cycle is defined as the ratio of the flare duration to the rotation period. Here the flare duration is determined when the envelope of the flare departs from the background emission. Among the objects with confirmed periodic pulsations, the duty cycle is $\sim$35\% for TVLM 513-46  \citep[][multiple pulses with the most luminous one at $X$ band lasting $\sim$8 min]{2007ApJ...663L..25H}, $\sim$12\% for LSR J1835+3259 \citep[][$\sim$20 min flare]{2008ApJ...684..644H}, $>$30\% for 2MASS J0036+1821 \citep[][$>$55 min flare]{2008ApJ...684..644H}, $\sim$0.8\% for 2MASS J0746+2000 \citep[][$\sim$1 min flare]{2009ApJ...695..310B}, $\sim$20\%-35\% for 2MASS J1314+1320 \citep[][multiple 10$^2$-10$^3$ s flares]{2015ApJ...799..192W}, and $\sim$7.8\%-12.6\% for 2MASS J1047+2124 \citep[][$\sim$500-800 s flare]{2015ApJ...808..189W}. Note that \citet{2002ApJ...572..503B} measured the duty cycle of TVLM 513-46 to be $\tau_{\rm flare} / t_{\rm on-source} \approx$13\% (see Table~\ref{tab1}). Apart from the diverse definition of the flare duration and duty cycle, the low time resolution of 5.5 min in \citet{2002ApJ...572..503B} may smooth the weak and short-timescale variability, resulting in a different value of duty cycle. For the case of 2MASS J1047+2124, \citet{2012ApJ...747L..22R} detected sporadic radio bursts, while \citet{2015ApJ...808..189W} and \citet{2020Sci...368..169A} measured a periodicity of $\sim$1.76 hr. One possibility is that the sensitivity of Arecibo observations in \citet{2012ApJ...747L..22R} is $\sim$0.2 mJy, lower than $\sim$0.09 mJy in \citet{2015ApJ...808..189W}. The majority of flares observed in \citet{2015ApJ...808..189W} and \citet{2020Sci...368..169A} are below $\sim$0.6 mJy (3$\sigma$ detection for Arecibo). Hence \citet{2012ApJ...747L..22R} would miss these weaker flares. Meanwhile, the lower sensitivity would shorten the flare timescales according to our definition. As the sensitivity of FAST observations is better than that in \citet{2015ApJ...808..189W} (discussed in Section~\ref{crafts}), we adopt the duty cycle of 2MASS J1047+2124 in \citet{2015ApJ...808..189W}. It is worth noting that the different duty cycles measured for one object can also be caused by the intrinsic variation of flares. Based on light travel-time arguments, the flare duration is related to the extension of the source region where ECM emission generates by multiplying the velocity of light \citep{2006A&ARv..13..229T}. A long flare timescale reflects an extended source region, and multiple pulses indicate several compact source regions. If the source region size changes, the flare duration and consequently the duty cycle would be changed. Considering the existing data are sparse to characterize the distribution of duty cycles, we simply assume a uniform distribution over the range from 0.8\% to 35\% in our simulation. That is, a random value of duty cycle in this range would be assigned to a modeled UCD each time, and the duty cycle variability would be taken into account in the performance of one thousand trials.

To describe the temporal evolution of the flare, we model the flare by exponentials with the same rise and decay time $\tau$, that is, $F(t) = F_0 e^{(t-t_0)/\tau}$ for $t < t_0$ and $F(t) = F_0 e^{(t_0-t)/\tau}$ for $t > t_0$, where $F_0$ is the peak flux density at $t_0$. By modeling the flares of TVLM 513-46 in \citet{2007ApJ...663L..25H}, LSR J1835+3259 in \citet{2008ApJ...684..644H} and 2MASS J0746+2000 in \citet{2009ApJ...695..310B}, we choose 1 percent of the peak flux density as the cutoff of the exponentials, and then $\tau = - 0.5 \times flare ~ duration/{\rm ln} (0.01)$. Thus for a flare, when the peak flux density, the rotation period and the duty cycle are given, the profile can be shaped. 

\section{Search for radio-flaring UCDs in CRAFTS} \label{potential}

CRAFTS, as introduced in Section~\ref{sec:intro}, is a large-scale commensal survey designed to observe extra-galactic H\,{\sc i}, Galactic H\,{\sc i}, pulsars and FRBs simultaneously using drift scans. Details on the survey can be referred to \citet{2018IMMag..19..112L}. Here we will discuss the possibility of utilizing the survey data for radio-flaring UCD search. 

\subsection{Basic characteristics of CRAFTS} \label{crafts}

FAST is located at (106.9$^\circ$~E, 25.7$^\circ$~N) with a declination range between -14.3$^\circ$ and 65.7$^\circ$, i.e. the maximum zenith angle ($ZA$) is 40$^\circ$ \citep{2018IMMag..19..112L}.
CRAFTS plans to carry out a two-pass survey with one-pass completion of $\sim$220 full days, covering $\sim$23,000 deg$^2$, approximately 60\% of the whole sky \citep{2018IMMag..19..112L}.
During the survey, FAST uses $L$-band Array of 19-beam (FLAN) receiver with a rotation of 23.4$^\circ$ to achieve a super-Nyquist sampling \citep{2018IMMag..19..112L}.
The spacing between two adjacent drift scans is 21.9$'$ in declination. FLAN covers a 400 MHz bandpass centered at 1250 MHz. When $ZA$ is below 26.4$^\circ$, a fully 300-m aperture is illuminated, and the beam size is approximately 3.7$'$ at 1050 MHz and 2.9$'$ at 1450 MHz. The antenna gain is $\sim$16.5 K/Jy, and the system temperature increases with $ZA$ from $\sim$18 K to $\sim$21 K at around 1250 MHz when $ZA < 26.4^\circ$. \citep{2019SCPMA..6259502J}.

\begin{figure}
 \includegraphics[width=\columnwidth]{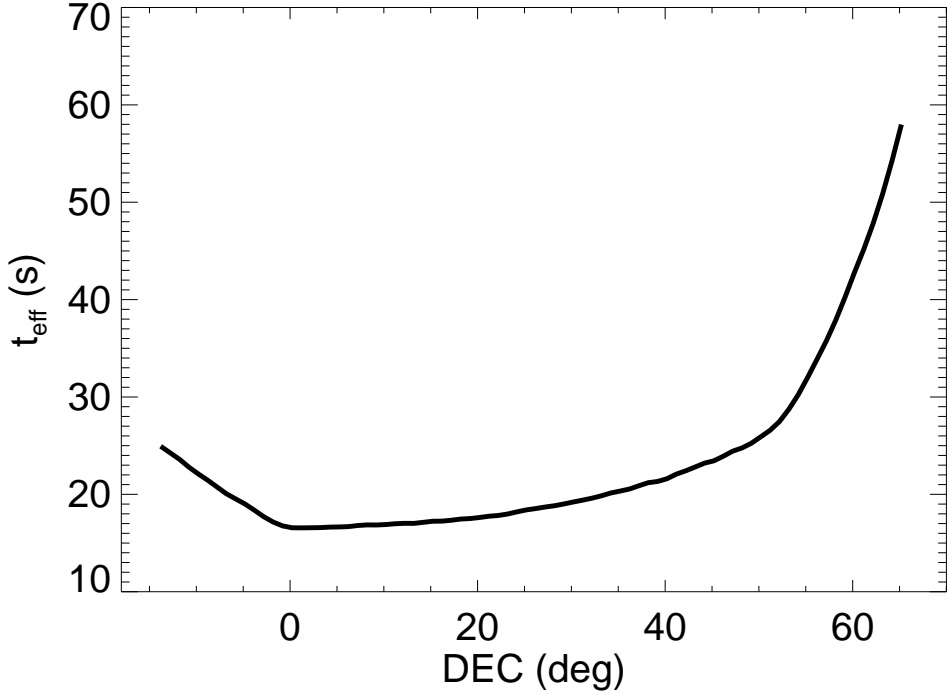}
 \caption{The effective integration time for CRAFTS one-pass drift scan as a function of declination at $\nu$ = 1450 MHz.   \label{fig1}}
\end{figure}

\citet{2019SCPMA..6259506Z} estimated the FLAN beam size ($\theta$), gain ($G$) and system temperature ($T_{\rm sys}$) during the drift-scan observations, based on the
FAST commissioning performance analysis from \citet{2019SCPMA..6259502J}. When a point source drifts through the FLAN at a speed of $v_{\rm s}$/arcmin s$^{-1}$=0.25 $\times$ cos(DEC), it would be seen by several beams. We use a 2D Gaussian to describe the beam response to a point source.  By doing a numerical experiment, we calculate the equivalent integration time, convolved with the Gaussian response curve, for each beam at 1450 MHz and add up all the time as the effective integration time ($t_{\rm eff}$). In Figure~\ref{fig1}, we show that $t_{\rm eff}$ ranges from $\sim$17 s at the equator to $\sim$60 s at the maximum declination of 65.7$^\circ$. The longer $t_{\rm eff}$ is due to the projection effect at high declination, and the increasing beam size due to partially-illuminated aperture at $ZA >$ 26.4$^\circ$. It is worth noting that there are gaps between beams and the total time drifting through a point source is approximately $2t_{\rm eff}$.

\begin{figure}
 \includegraphics[width=\columnwidth]{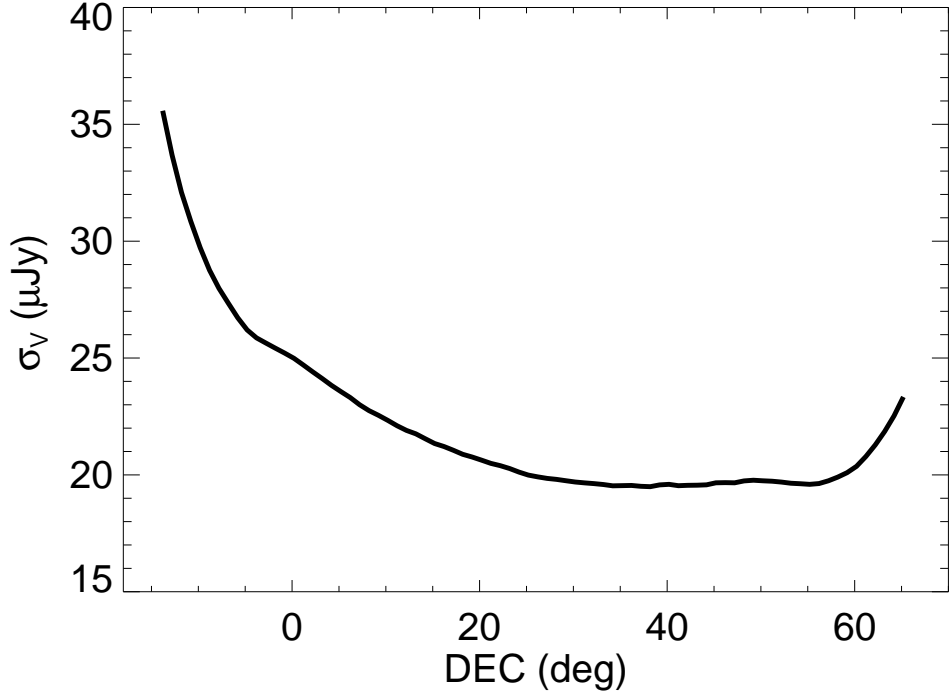}
 \caption{The rms noise in Stokes $V$ for CRAFTS one-pass drift scan as a function of declination. 320 MHz-bandwidth is used. The integration time is shown in Figure~\ref{fig1}. \label{fig2}}
\end{figure}

In FLAN, dual-linear polarizations are available for each beam. The rms noise in a single polarization can be estimated by 
\begin{equation}
\sigma = \frac{T_{\rm sys}}{G} \frac{1}{\sqrt{t_{\rm eff} \Delta \nu}},
\label{sensitivity}
\end{equation}
$\Delta \nu$ is the bandwidth, and we use the full range 400 MHz, deducting 20\% bandwidth loss due to RFI based on experience. The rms noise in Stokes $V$ ($\sigma_V$) is $\sqrt{2}$ times larger than that in a single polarization (see Figure~\ref{fig2}). The system temperature increases as a function of $ZA$ and the gain drops off when $ZA >$ 26.4$^\circ$ due to the partially-illuminated aperture. The variation of $\sigma_V$ with the declination results from the competition of the effects of different parameters. 

Stokes $V$ suffers little confusion, and for linear polarization feeds, Stokes $V$ is independent of gain fluctuation of the system when the two polarization paths are well isolated \citep{2001PASP..113.1243H}. Thus the measurement of Stokes $V$ is a good method to search for flares. The processing of Stokes V data refers to the RHSTK package \footnote{Heiles, Carl; Robishaw, Tim; Kepley, Amanda; and Furea Kiuchi. 2012.
All-Stokes Single Dish Data with the RHSTK (Robishaw/Heiles SToKes) Software Package.}. The gain and phase response of the system can be calibrated by the noise diode (CAL). For simplicity, the instrumental contribution to polarimetry is described using a unitary matrix, in which only zeroth-order effects are considered \citep{2001PASP..113.1274H}. In this case the uncertainty of Stokes $V$ due to the leakage of other Stokes parameters, mainly Stokes $I$, can be up to $\sim$10\%. Thus only the detection of radio emission with circular polarization $>$10\% is reliable. It is acceptable for flaring UCD search, as ECM emission is highly circularly polarized. Higher precision polarimetry is available when more corrections are made to remove the leakage of other Stokes parameters to Stokes V. Yet it is beyond the scope of this paper.

\subsection{Mock UCD catalog generation} \label{mcs}

In this section, we describe the Monte Carlo simulation for generating mock catalog, and discuss how the flaring radio luminosity and distance are assigned to each source for FAST detection assessment.

\begin{figure}
 \includegraphics[width=\columnwidth]{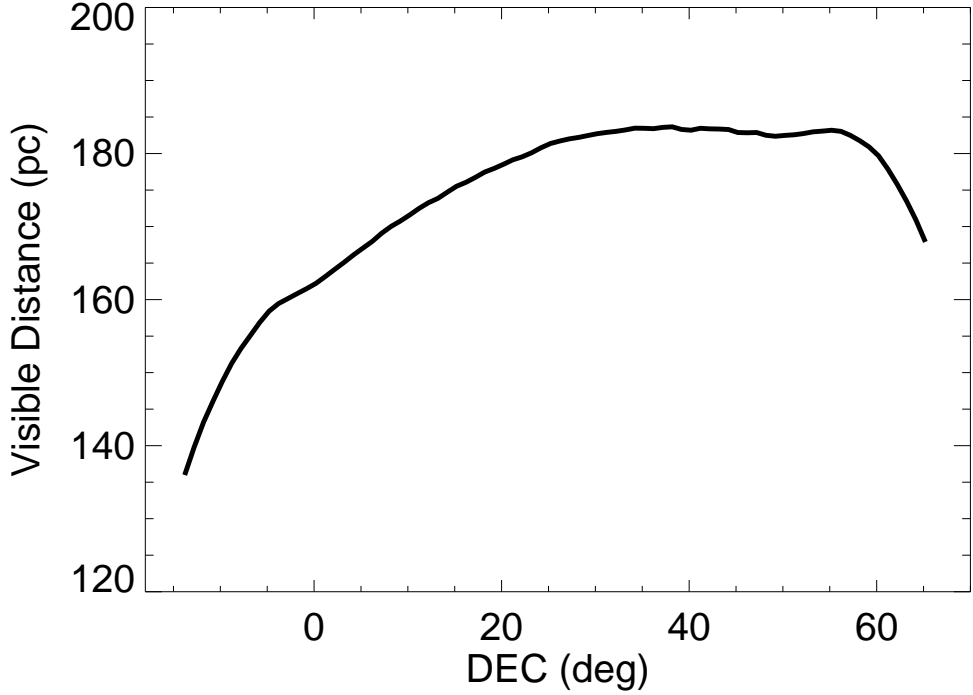}
 \caption{The visible distance of DENIS J1048-3956 flare with $\nu L\nu$ = 4.92 $\times$ 10$^{24}$ erg s$^{-1}$
 for CRAFTS one-pass drift scan
 as a function of declination at $\nu$ = 1250 MHz.
 \label{fig3}}
\end{figure}

In Table~\ref{tab1}, DENIS J1048-3956 is found to have the most luminous flare with $\nu L\nu$ = 4.92 $\times$ 10$^{24}$ erg s$^{-1}$. This luminosity can be detected with $\ge 5\sigma_V$ by CRAFTS at 1250 MHz at a distance up to $\sim$180 pc (see Figure~\ref{fig3}), though this object itself is out of FAST sky coverage. In comparison, the scale height of low-mass stars in the Galactic thin disk is $\sim$300 pc \citep{2008ApJ...673..864J, 2009ApJ...695.1591P, 2010AJ....139.2679B}. To count the total number of all (radio emitting or not) UCDs within this distance, we consider the space density of UCDs. The local density of UCDs is calculated based on the compiled results from a number of volume-limited ($<$ 25 pc) surveys. The space density is $\sim$11.1 $\times$ 10$^{-3}$ pc$^{-3}$ for M7--L4 dwarfs \citep{2019ApJ...883..205B}, $\sim$3.4 $\times$ 10$^{-3}$ pc$^{-3}$ for L5--T5.5 dwarfs \citep{2010A&A...522A.112R}, $>$ 14.6$ \times $10$^{-3}$ pc$^{-3}$ for T6 and later dwarfs with $T_{\rm eff} <$ 1050 K \citep{2019ApJS..240...19K}. These numbers have been accounted for the completeness. We follow the similar treatment in \citet{2017ApJ...845...66R} to sum up these objects of different spectral types to obtain a total space density of $>$ 2.9 $\times$ 10$^{-2}$ pc$^{-3}$ for UCDs.

Here we only consider the thin disk contribution for UCD space density
\begin{equation}
\rho=\rho_0 e^{-\frac{|Z|-Z_\odot}{H}}, 
\label{eq1}
\end{equation}
where $\rho_0$ is the local density, $Z$ is the vertical distance from the plane, and $H$ is the scale height. We adopt $\rho_0 = 2.9 \times 10^{-2}$ pc$^{-3}$ as discussed above, $Z_\odot$ = 15 pc as the Galactic height of the Sun \citep{1995ApJ...444..874C, 1997A&A...324...65N, 1997MNRAS.288..365B}, and $H = $ 300 pc as the scale height for UCDs \citep{2008ApJ...673..864J, 2009ApJ...695.1591P, 2010AJ....139.2679B}. The influence of different scale heights to our estimates will be discussed later in Section~\ref{scaleheight}. 
The Galactic height ($Z$) of each point is calculated by the equation given in \citet{2010AJ....139.2679B}:
\begin{equation}
Z=Z_{\odot}+d\sin(b-\arctan(Z_{\odot} / R_{\odot})),
\end{equation}
where $d$ is the distance, $b$ is the Galactic latitude. $R_\odot$ = 8.5 kpc is taken as the Galactic radius of the Sun \citep{1986MNRAS.221.1023K}.
We note that the scale length of the thin disk is $\sim$3 kpc, significantly larger than the distance of a few hundred pc we consider here for CRAFTS (as shown in Figure~\ref{fig3}), so the density variation along the radial direction is ignored in our calculation.

We sample the solar ambient space into small volume elements by each $\alpha$ (RA), $\delta$ (DEC), $d$ bin, $\Delta \alpha$ = 1$^\circ$, $\Delta \delta$ = 1$^\circ$, $\Delta d$ = 5 pc, marked with ($\alpha$, $\delta$, $d$). The UCD density $\rho (\alpha, \delta, d)$ of each volume element is estimated based on Equation \ref{eq1}, after converting the center coordinates of the volume element from the equatorial coordinates ($\alpha$, $\delta$) to the galactic coordinates ($l$, $b$) and further to the cylindrical coordinates ($R$, $Z$). An azimuthal symmetry of the Galactic disk \citep{2008ApJ...673..864J} is assumed.
Multiplied by the individual volume size, we obtain the UCD number in each volume element $N (\alpha, \delta, d)$, and then integrate over $\alpha$ to obtain $N (\delta, d)$. In a given $\delta$, $d$ bin, we generate $N (\delta, d)$ uniformly distributed random numbers, and then sum over $d$ up to the farthest visible distance related to $\delta$, denoted by $N (\delta)$. These artificial sources are given the radio luminosities following the empirical LF, i.e. $d N/d L \propto L^{-1.96}$ with $L_{\rm min}$ = 4.2 $\times$ 10$^{23}$ erg s$^{-1}$ (Section~\ref{lf}), the rotation periods following the lognormal distribution with $\mu_0$ = 1.41 and $\sigma_0$ = 0.48 (Section~\ref{sec:duty_cycle}), and the duty cycles uniformly distributed from 0.8\% to 35\% (Section~\ref{sec:duty_cycle}). 

\begin{figure}
 \includegraphics[width=\columnwidth]{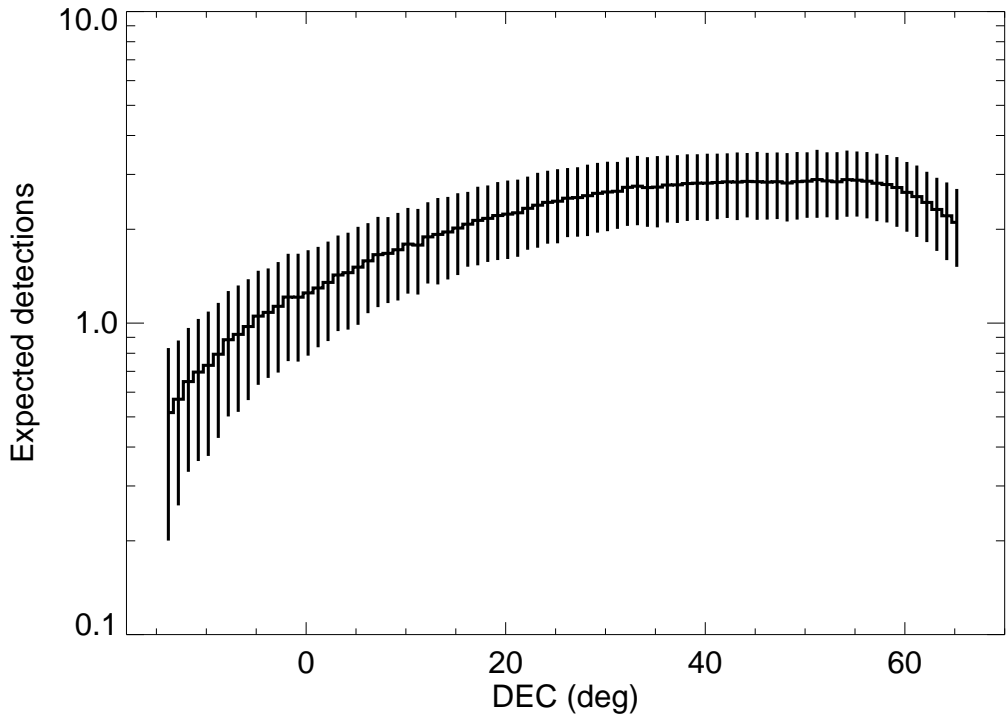}
 \caption{The average number of the flaring UCDs that are expected to be detected during the two-pass drift scans with vertical standard deviation bars obtained from one thousand trials. The DEC bin is 1$^\circ$.
\label{fig4}}
\end{figure}

We randomly select 3\% of these objects (Section~\ref{rate}), and calculate their flux densities at 1250 MHz (assuming $\nu L_{\nu}$ is flat). As discussed in Section~\ref{sec:duty_cycle}, the temporal evolution of the flare can be determined when the peak flux density ($F_0$), the rotation period ($P$) and the duty cycle are specified. Here we assume the flaring emission is 100\% circularly polarized \citep{2007ApJ...663L..25H, 2008ApJ...684..644H, 2008AA...487..317A, 2009ApJ...695..310B, 2015ApJ...799..192W, 2015ApJ...808..189W, 2016ApJ...821L..21R}. Meanwhile, for the object with given $\delta$, the effective integration time $t_{\rm eff}$ and the corresponding rms noise $\sigma_V$ are known (see Figure~\ref{fig2} and Figure \ref{fig3}). In this case, 
we can derive how long the flare lasts with the flux density $\ge 5\sigma_V$, which is set as the detection limit. This time duration is denoted by $t_{\rm d}$. Here $\sigma_V$ is calculated with $t_{\rm eff}$, while the total drifting time is approximately $2t_{\rm eff}$ as noted in Section~\ref{crafts}. That means when FAST drifts through a source with $2t_{\rm eff}$, the source should keep in flaring state brighter than $5\sigma_V$. Considering the marginal $5\sigma_V$ detection, $2t_{\rm eff}$ can be half above and half below $5\sigma_V$ flux limit. Therefore, the probability of each source being detected is $p=t_{\rm d}/P$. The sources whose $F_0 < 5\sigma_V$ or $t_{\rm d} < 2 t_{\rm eff}$ are excluded. During the two-pass survey, each point source would be observed twice. The flux variation in Stokes $V$ caused by the flare can be potentially identified, as Stokes $V$ suffers little confusion and gain fluctuation of the system. We calculate the probability of each source being detected once in two passes, that is, $2p(1-p)$, and add up all the probabilities to obtain the expected value. The average of the expected detections of flaring UCDs in CRAFTS and the corresponding standard deviation at each declination from a thousand Monte Carlo realizations are shown in Figure~\ref{fig4}. The total expected detections are $\sim$170, based on $d N/d L \propto L^{-1.96}$ with $L_{\rm min}$ = 4.2 $\times$ 10$^{23}$ erg s$^{-1}$. Considering the uncertainty of the LF power-law index (Section~\ref{lf}), we also perform the simulations with $\alpha$ = 1.51 and $\alpha$ = 2.41. The expected detections are $\sim$500 and $\sim$80, respectively.

\section{Discussion} \label{discuss}

\subsection{Alternative assumptions for the simulation}\label{sec:parameter}

In the Monte Carlo simulation, we assume a thin-disk dominated space density distribution of UCDs with a scale height of 300 pc, a 3\% flaring UCD detection rate, a radio LF of the flares with a power-law index of 1.96, a flat $\nu L_\nu$ between 1-10 GHz, and a uniform distribution of flare duty cycle from 0.8\% to 35\%. Under these assumptions, we estimate $\sim$170 flaring UCDs will be detected in CRAFTS. The relevant parameters are chosen mainly to yield a conservative estimate.
In the following, we will discuss the uncertainties of these parameters and their impact on the results.

\subsubsection{The scale height} \label{scaleheight}
When we estimate the inhomogeneous spatial distribution of UCDs, we considered the thin disk contribution with a scale height of 300 pc for low-mass stars (Section~\ref{mcs}). This scale height for
brown dwarfs was also measured, but the value suffers from a large uncertainty owing to
the faintness of brown dwarfs and the resultant small number statistics. \citet{2005ApJ...622..319P} used six dwarfs with spectral type M4 and later (including two L dwarf candidates) to derive a Galactic disk scale height of 400 $\pm$ 100 pc for M and L dwarfs. \citet{2005ApJ...631L.159R} estimated the scale height of 350 $\pm$ 50 pc based on 28 L and T dwarfs, and \citet{2011ApJ...739...83R} drew the value of 290 $\pm$ 25 (random) $\pm$ 31 (systematic) from 17 newly-discovered UCDs later than M8. \citet{2019ApJ...870..118S} detected 3665 L dwarfs and gave a good fit for the scale height of 380 pc. 

\begin{figure}
 \includegraphics[width=\columnwidth]{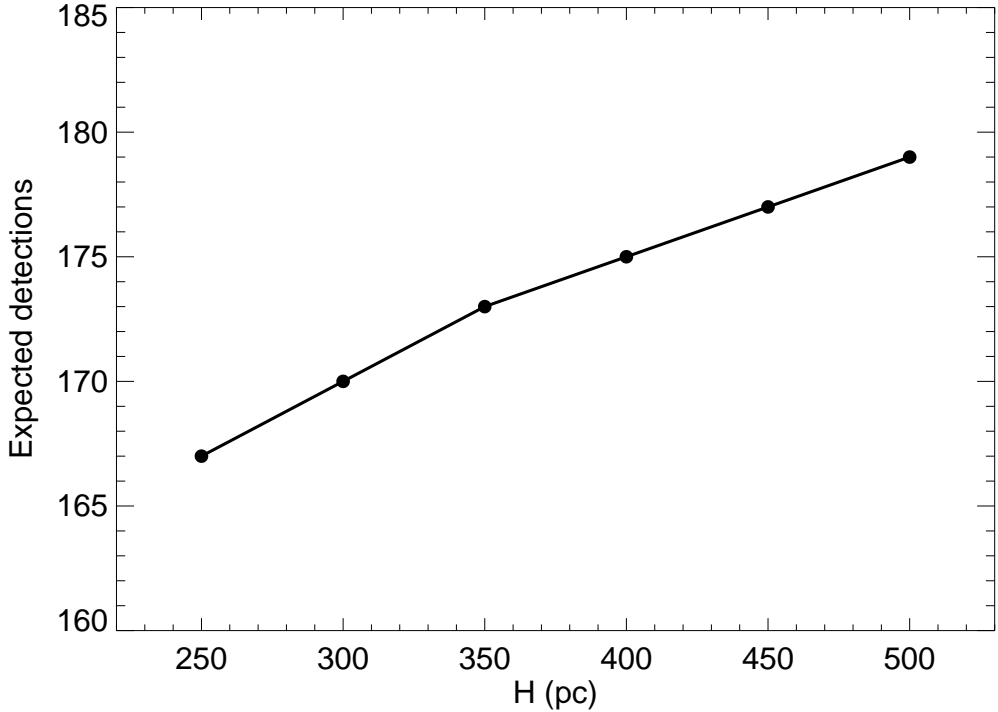}
 \caption{The expected detection number of flaring UCDs during CRAFTS varies with the scale height $H$ of the Galactic thin disk. A larger scale height indicates more UCDs in CRAFTS sky coverage.
 \label{fig5}}
\end{figure}

A scale height larger than the 300 pc adopted here will not significantly affect our results because the maximum visible distance of UCD flares in CRAFTS is only about 180 pc (Figure \ref{fig3}). 
For the purpose of comparison, we conduct the simulations with different values of the scale height. The results are shown in Figure~\ref{fig5}.

\subsubsection{The detection rate}
We derived a detection rate of radio-flaring UCDs to be $\ge$3\% from a series of unbiased, targeted surveys conducted by both interferometers and single dish telescope in Section~\ref{rate}, and adopted the value of 3\% in the simulation. Another estimate is referring to the survey detection statistics of Arecibo only, since FAST and Arecibo are both single dish telescopes, and their instrumental properties are similar. They are sensitive to rapidly varying emission, but impractical to detect weak, quiescent emission due to the large beams and the related confusion limitations. In the two surveys of 51 not-previously detected sources, two flaring dwarfs were seen by Arecibo \citep{2013ApJ...773...18R, 2016ApJ...830...85R}, yielding a detection rate of $\sim$4\%. With this value, the expected detection number is 226. 

\subsubsection{The spectral index} \label{spectralindex}
Assuming the flux density spectrum of the flare is $S_{\nu} \propto \nu^{\beta}$, the spectral index $\beta$ would affect the simulation significantly. The measurements of spectral indices are limited, as the multi-band simultaneous radio observations are unusual. \citet{2017ApJ...834..117W} observed WISEP J1122+2550 using VLA at 4-8 GHz. They generated two images by splitting the 4 GHz bandwidth into halves with central frequencies of 5.00 and 7.00 GHz, and implied $\beta$ = -1.5 $\pm$ 0.3. It can be taken as the spectral index of flaring emission, as they considered there may be no quiescent emission in this dwarf. \citet{2018ApJS..237...25K} used VLA to observe four radio-emitting brown dwarfs at 8-12 GHz and one at 12-18 GHz. For those observed at 8-12 GHz, we adopt pulse flux densities at sub-bands 8-10 and 10-12 GHz with central frequencies of 9.00 and 11.00 GHz, and estimate $\beta$ = -1.7 $\pm$ 1.3 for SIMP J0136+0933 and $\beta$ = -1.4 $\pm$ 0.8 for 2MASS 1043+2225 (see Table 7 in \citet{2018ApJS..237...25K}, values in ``All Pulses" column are adopted for 2MASS 1043+2225). For 2MASS 1237+6526, there are three pulses (see Table 5 in \citet{2018ApJS..237...25K}), and $\beta$ = 2.0 $\pm$ 1.7, 0.7 $\pm$ 0.4 and 1.5 $\pm$ 0.9, respectively. For SDSS 0423-0414, there are two right-circularly polarized pulses, labeled as Pulse R1 and R2, and four left-circularly polarized pulses, labeled as Pulse L1, L2, L3 and L4 (see Table 8 in \citet{2018ApJS..237...25K}). For Pulse L2 and L4, pulses are detectable at 8-10 GHz, but not at 10-12 GHz. It may be a hint of ECM emission frequency cutoff. Thus we calculate spectral indices for another four pulses and $\beta$ = -0.4 $\pm$ 1.1, -2.7 $\pm$ 1.3, -2.9 $\pm$ 1.1 and -2.2 $\pm$ 0.9, respectively. In addition, 2MASS 1047+2124 was observed at 12-18 GHz, and six pulses were detected. The 6 GHz bandwidth is divided into four sub-bands. These pulses were detectable in one sub-band at most, except Pulse 5 (see Table 6 in \citet{2018ApJS..237...25K}). For this pulse, considering the flux densities of 93.7 $\pm$ 24.0 $\mu$Jy at 15.75 GHz and 91.5 $\pm$ 28.7 $\mu$Jy at 17.25 GHz, the spectral index is $\beta$ = -0.3 $\pm$ 4.4.

On the other hand, from the perspective of ECM theory, electrons are accelerated in the magnetic field-aligned electric fields. Under the mirror effect of the converging magnetic field-lines, the energy gained in the electric fields is transferred from parallel to perpendicular energy (with respect to the magnetic field). The resulting electron-velocity distribution is unstable. Its positive gradient in the perpendicular velocity leads to the exponential growth of the radiation. The intensity of ECM emission depends on the magnetic field strength and structure, the electron density and energy distribution, the electron-distribution function in velocity space and the radiation source region size \citep{2006A&ARv..13..229T}. Observed ECM emissions at different frequencies are considered to originate from source regions at different heights over the surface. Their flux densities are not predictable, as not all aforementioned parameters in localized source regions are known.

In this work we assume $\beta$ = -1.0, that is, $\nu L_{\nu}$ is flat, and consider other values ranging from -2.5 to 0 with a step of 0.5 to investigate the effect of $\beta$ on the simulation. This assumption is not unreasonable. In the above-mentioned seven cases, the spectral indices of WISEP J1122+2550, SIMP J0136+0933 and 2MASS 1043+2225 are between -1.0 and -2.0, while the median value of SDSS 0423-0414 is steeper. If taking into account the uncertainties, $\beta$ = -1.0 is applicable to these objects and also 2MASS 1047+2124. Moreover, \citet{2020ApJ...903L..33V} discovered BDR J1750+3809 at around 144 MHz with a radio spectral luminosity of $\sim$5 $\times$ 10$^{15}$ erg s$^{-1}$ Hz$^{-1}$. The radio luminosity is $\nu L_{\nu} \approx$ 7.2 $\times$ 10$^{23}$ erg s$^{-1}$, consistent with the values listed in Table~\ref{tab1}. It is worth noting that \citet{2018ApJS..237...25K} found pulses seem more intermittent at high frequencies. They proposed one possibility that while large-scale fields are necessary to drive auroral currents to produce radio flaring emissions, small-scale fields may emerge near the surface of the object, causing rapid variation of pulses in both time and frequency. This phenomenon begins to appear around 10 GHz for SDSS 0423-0414 and seems more obvious above 12 GHz in the case of 2MASS 1047+2124. As the origin of this phenomenon is unclear yet, we note that it is safer to apply the assumption of flat $\nu L_{\nu}$ to radio emissions below 10 GHz. 

\begin{figure}
 \includegraphics[width=\columnwidth]{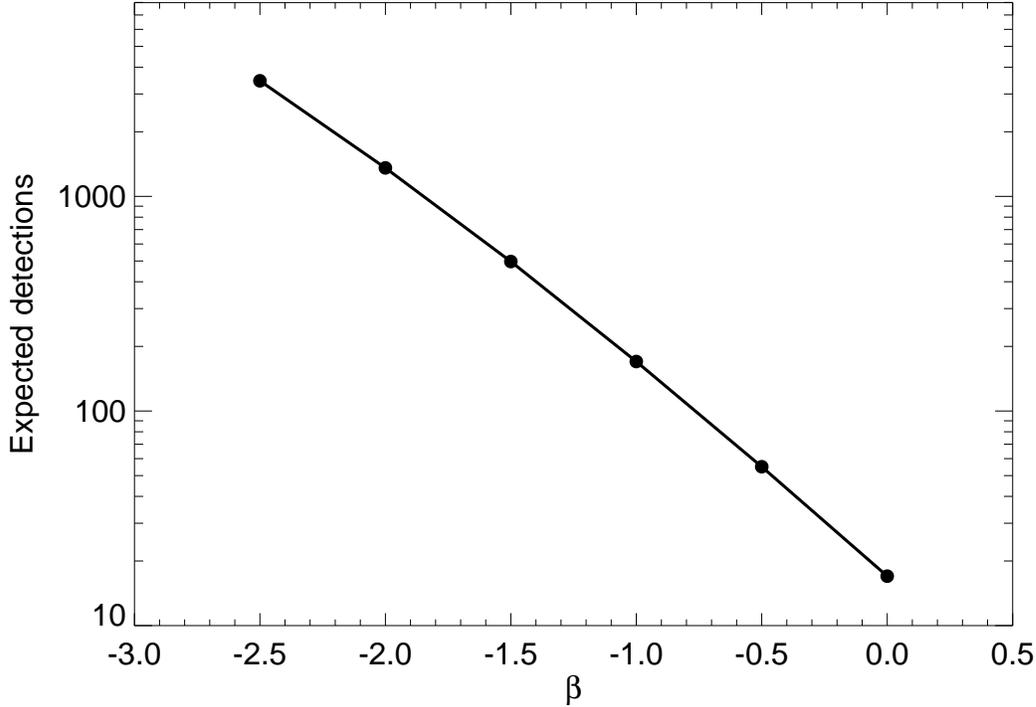}
 \caption{The expected detection number of flaring UCDs during CRAFTS varies with the spectral index $\beta$ of the flaring emission. A more negative value of $\beta$ implies a stronger $L$-band flux. The visible distance of UCD flares in CRAFTS thus extends further, and the detection number increases as a result.
 \label{fig6}}
\end{figure}

The simulation results with different values of $\beta$ are shown in Figure~\ref{fig6}. When $\beta$ = 0, $<$ 20 radio-flaring UCDs would be detected. A more negative value of $\beta$ yields more expected detections, as the inferred $L$-band flux would be higher. If a more extreme $\beta$ = -2.5 is considered, the expected detection number would increase dramatically to $\sim$3460, one order of magnitude larger than that predicted from $\beta$ = -1.0. Thus the future survey results would help constrain the average spectral index.

\subsubsection{The LF index}
When we derive the LF index $\alpha$ in Section~\ref{lf}, we follow \citet{doi:10.1137/070710111} to normalize the constant $k$ by integrating to infinity. In reality, the probability would be zero at the luminosity higher than the most luminous event ($L_{\rm max}$) \citep{2015ApJ...814...19A}, and thus the normalization should be written as $k = (\alpha-1)/(L_{\rm min}^{-\alpha+1}-L_{\rm max}^{-\alpha+1})$. The existence of $L_{\rm max}$ makes the LF flatter. Without further knowledge of $L_{\rm max}$ from limited surveys on UCD flares, here we consider an extreme case that it is the currently observed maximal luminosity, that is, $L_{\rm max}$ = 4.92 $\times$ 10$^{24}$ erg s$^{-1}$. By maximizing Equation~\ref{mle}, we obtain $\alpha$ = 1.37. The expected detection number of flaring UCDs is $\sim$100. For more comparison, we also run the simulation with $L_{\rm max}$ = 9.84 $\times$ 10$^{24}$ erg s$^{-1}$ and the corresponding $\alpha$ = 1.69. The result is $\sim$130. 

\subsection{Confirmation of flaring UCD candidates }\label{follow-up}

When we use the Stokes $V$ flux difference between two-pass observations to search for radio-flaring UCDs, the sample will be contaminated by other radio sources with variability. Apart from UCDs and planets, the known radio transients emitting the continuum with high circular polarization are stars (earlier than M7) and pulsars. Cross-matching the archival information about that source on the sky would be helpful for early identification. By this means \citet{2021MNRAS.502.5438P} identified 33 radio stars in the Rapid ASKAP Continuum Survey (RACS) and \citet{2021NatAs...5.1233C} reported radio detection of 19 M dwarfs in LoTSS with $\approx$ 20\% of the Northern sky processed. The catalog of brown dwarfs is far from complete in solar neighborhood so far, yet it will be greatly improved in future infrared surveys, e.g. Euclid, SPHEREx and Roman. Moreover, we need the follow-up observation for confirmation and detailed studies, especially for those distant and faint objects, which could be beyond the scope of other radio telescopes.

\begin{figure}
 \includegraphics[width=\columnwidth]{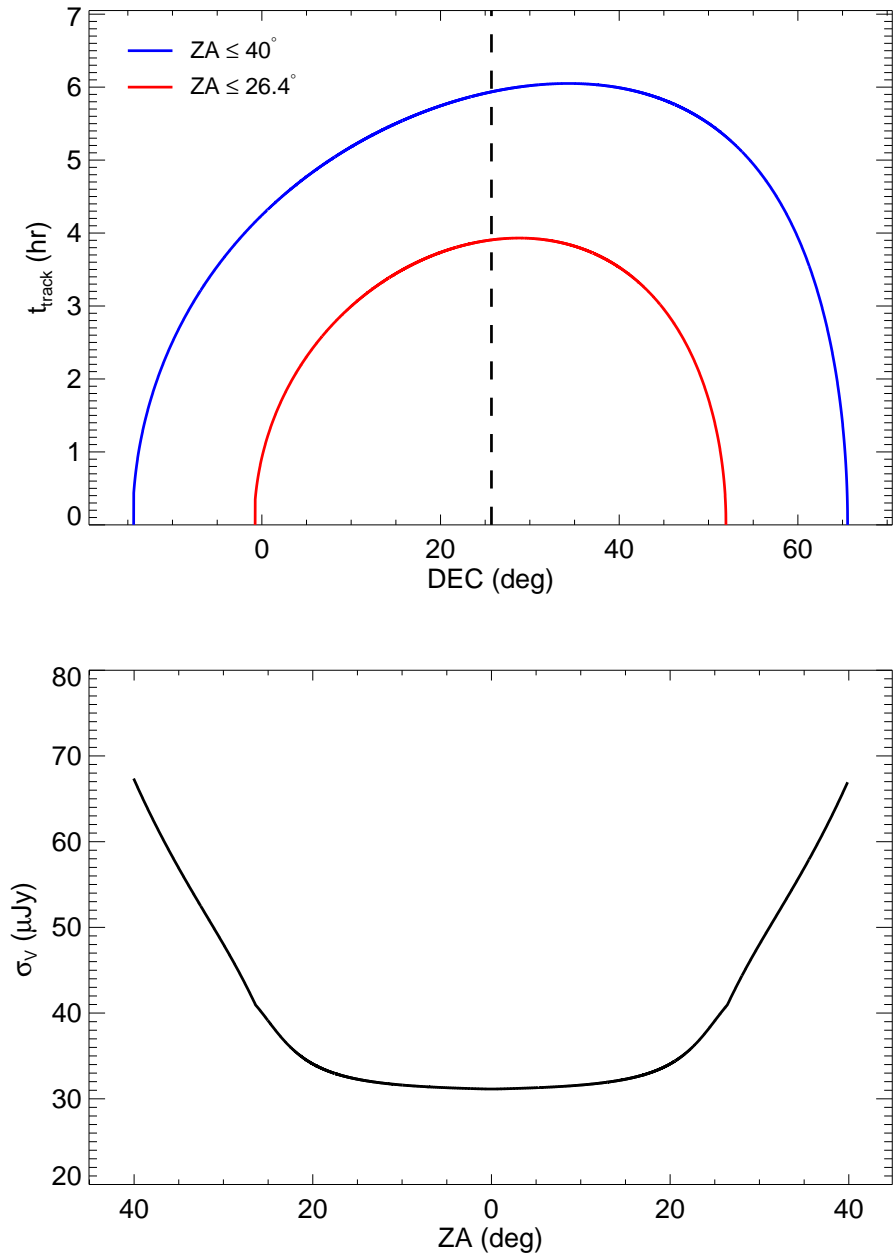}
 \caption{Top: Maximum tracking time vs. target declination when $ZA$ is below 40$^\circ$ (blue) and 26.4$^\circ$ (red). The dashed line marks the latitude of FAST. This plot is from \citet{2020RAA....20...64J}. Bottom: The rms noise in Stokes $V$ when tracking as a function of $ZA$ with 10 s time resolution for reference.
 \label{fig7}}
\end{figure}

The strategy of selecting flaring UCD candidates in CRAFTS mainly depends on
two factors: the available tracking time of the target and the corresponding FAST sensitivity. The top plot in Figure~\ref{fig7} shows the maximum tracking time for a target depending on its declination. The sources with the declination between 3.4$^\circ$ and 49.2$^\circ$ can be observed continuously for at least 2 hr when keeping $ZA \leq$ 26.4$^\circ$. If $ZA >$ 26.4$^\circ$ (with reduced sensitivity) is acceptable, the 2 hr long observation would be available to the sources with the declination from -11.7$^\circ$ to 64.4$^\circ$.
The bottom plot in Figure~\ref{fig7} shows the sensitivity variation of the telescope in Stokes $V$ with $ZA$ when FAST works in tracking mode. The sensitivity in a single polarization is calculated by Equation~\ref{sensitivity} with the reported gain and system temperature at different frequencies in \citet{2020RAA....20...64J}, and the average is adopted over the frequency. The sensitivity in Stokes $V$ is $\sqrt{2}$ times larger than this value. When tracking the sources with the declination below -0.7$^\circ$ or above 52.1$^\circ$, the minimum $ZA$ is 26.4$^\circ$. That means the best sensitivity in Stokes $V$ (with 10 s time resolution for reference in Figure~\ref{fig7}) is $\sim$41.0 $\mu$Jy, reduced by approximately 31\% compared to the value of $\sim$31.2 $\mu$Jy near $ZA$ = 0$^\circ$, and the sensitivity decreases quickly with increasing $ZA$. For these sources, a more sophisticated observation scheme should be designed to compensate the sensitivity requirement and the continuous flux monitoring duration.

\subsection{Implication of flaring UCD search in $L$-band survey}
Since radio-active UCDs were discovered at $\sim$5 GHz and $\sim$8 GHz, as summarized in Section ~\ref{surveys}, lower frequency observations have been attempted to explore the nature of the radio emission at low frequencies. \citet{2006ApJ...637..518O} conducted VLA observations of TVLM 513-46 at 8.4, 4.8 and 1.4 GHz. They detected persistent radio emission at all frequencies, yet no flares. Later \citet{2009ApJ...700.1750O} observed another two radio sources 2MASS J0523-1403 and LHS 3003 at these three frequencies, and no radio emission was detected. \citet{2011ApJ...741...27M} detected quiescent emission of $\sim$1 mJy at  1.43, 4.86, 8.46 and 22.5 GHz from 2MASS J1314+1320 with VLA. \citet{2017MNRAS.465.1995M} used the upgraded VLA to observe six radio-emitting UCDs at $S$ (2-4 GHz) and $C$ (4-7 GHz) band. Quiescent radio emissions were detected from five sources at both bands, while the flaring emission was only detected from LSR J1835+3259 at $S$ band. This was the first time that a flare below $\sim$4 GHz was detected from a UCD. Moreover, \citet{2019MNRAS.483..614Z} presented the results of nine UCDs observed by GMRT at $\sim$610 and $\sim$1300 MHz. Only quiescent emissions were detected from 2MASS J1314+1320 and 2MASS J0746+2000 at these two frequencies. Hundreds of MHz observations of UCDs were also carried out. Unfortunately, neither VLA observations of two UCDs at 325 MHz \citep{2011AJ....142..189J} nor LOFAR mini-survey of three UCDs at around 140 MHz \citep{2016MNRAS.463.2202B} detected any radio emission. Though non-detection from target objects, \citet{2020ApJ...903L..33V} discovered a circularly polarized radio source at around 144 MHz in LoTSS. Follow-up near-infrared observations confirmed that it is a T6.5 brown dwarf at a distance of 65 pc. This is the first radio-discovered brown dwarf. In view of the high brightness temperature and circular fraction, the detected radio emission is interpreted as ECM emission, and the field strength of this object is inferred to be at least 25 Gauss.

The effort on low frequency exploration, though appreciated, is far from enough. As the flaring emission at $\sim$1 GHz has not been detected from any UCD yet, our understanding of the nature of such emission is limited. ECM emission is radiated mostly at the fundamental and the second harmonic of electron-cyclotron frequency, $\nu_c = eB / 2 \pi m_e c \approx 2.8 \times B$ [MHz]. Thus the observed flaring emission can be used to measure the local magnetic field, and 1.4 GHz emission corresponds to a magnetic field of 250--500 Gauss. It could be the case: (1) the source region with this field strength is located at a certain height over the surface of an object where the field strength is stronger and (2) this is the maximum surface field strength of an object. As mentioned in Section~\ref{sec:intro}, \citet{2009Natur.457..167C} proposed an energy flux scaling law working from rapidly rotating low-mass stars to solar system planets. Based on this scaling law, \citet{2009ApJ...697..373R} expressed the field strength in terms of mass $M$, luminosity $L$ and radius $R$, that is, $B = 4.8 \times (ML^2/R^7)^{1/6}$ [kG] with all parameters normalized with solar values.  Further \citet{2010A&A...522A..13R} calculated the evolution of average magnetic fields in brown dwarfs and exoplanets. They claimed that magnetic fields in brown dwarfs and planets weaken over time as they become fainter, losing luminosity which is the power source of magnetic field generation. According to the calculation \citep[see Figure~1 in][]{2010A&A...522A..13R}, a $5M_{\rm J}$ planet is predicted to have a polar dipole field strength of $\sim$500 Gauss during the first few Million years, and a brown dwarf with $M = 25 M_{\rm J}$ can maintain an average surface field of $\sim$250 Gauss after 10 Gyr. Objects with masses between them would own surface magnetic fields of 250--500 Gauss for some time during their lifetime. Flaring UCD search in $L$-band survey is expected to find such objects including exoplanets. Alternatively, \citet{2018ApJS..237...25K} found that late L dwarfs marginally follow the energy flux scaling law, and T dwarfs clearly depart from it based on studies of five targets. Radio-flaring UCDs and planets with diverse masses, luminosities and radii discovered in the survey can be used to test this scaling law. For instance, assuming a $\sim 10M_{\rm J}$ ($0.01M_{\odot}$) planet with a radius of $\sim 1R_{\rm J}$ ( $0.1R_{\odot}$) was discovered in the survey, it meant the field strength of this object was at least $\sim$500 Gauss, considering the fundamental electron-cyclotron frequency. If its luminosity was much smaller than $L = \sqrt{(B/4.8)^6 R^7/M} \approx 3.6 \times 10^{-6} L_{\odot}$, it could be considered to deviate from the energy flux scaling law. Infrared data are needed to derive these physical parameters. Note that this luminosity-driven dynamo is tenable for dipole-dominated fields in fast rotators whose rotation rate is above some threshold value and field strength is independent of rotation rate. For objects that rotate not fast enough for dynamo saturation, their field strengths and topologies depend on rotation \citep{2006A&A...446.1027C, 2006ApJ...638..336D, 2008ApJ...676.1262B}. New sample discovered in the survey with different rotation rates would be helpful to investigate the influence of rotation on the dynamos.

\section{Conclusions} \label{conclusion}

In this paper, we evaluate the capabilities of FAST to detect radio-flaring UCDs during the two-pass drift-scan survey CRAFTS. In the previous surveys,
the overall detection rate of radio-active UCDs is $\sim$10\%. Compiling the data from a number of unbiased, targeted radio surveys, we have estimated a flaring UCD detection rate of $\ge$3\%, and constructed a flare luminosity function $d N/d L \propto L^{-1.96 \pm 0.45}$ (here $L = \nu L_{\nu}$).

The most luminous radio flare detected in these surveys is from DENIS J1048-3956 with $\nu L _{\nu}$ = 4.92$\times$10$^{24}$ erg s$^{-1}$. The similar flaring events can be seen by CRAFTS up to $\sim$180 pc at 1250 MHz.
Taking into account the spatial distribution of these dwarfs with a scale height of 300 pc, we have performed a Monte Carlo simulation to generate mock catalog with distances, radio luminosities, rotation periods and flare duty cycles, which contains $\sim$3.26 $\times$ 10$^5$ UCDs in the CRAFTS sky coverage. Adopting the aforementioned $\sim$3\% flaring UCD fraction, we have estimated that $\sim$170 dwarfs are expected to be detected as transients, assuming 100\% circularly polarized emission. Radio-flaring UCDs discovered in the survey would gain our understanding of UCD behavior at $L$ band and dynamos.

While a giant single dish telescope such as FAST has the advantage to detect faint radio flares at low frequencies, the radio interferometers can detect slowly varying radio emission that is not accessible to the single dish due to its large beam size and other practical limitations. Together with the forthcoming SKA, these two complementary, world-class telescopes are expected to shed light on the magnetic fields and dynamos of UCDs. 

\normalem
\begin{acknowledgements}
We thank the referee for useful comments to improve the manuscript. We thank Carl Heiles, Tao-Chung Ching, and J. Davy Kirkpatrick for helpful discussions. This work is supported by the National Natural Science Foundation of China grant No. 11988101, No. 11725313, No. 11690024, No. 11703039, by the International Partnership Program of Chinese Academy of Sciences grant No. 114A11KYSB20160008, and by the CAS Strategic Priority Research No. XDB23000000. FAST is operated by the Open Project Program of the Key Laboratory of FAST, Chinese Academy of Sciences.
\end{acknowledgements}
  
\bibliographystyle{raa}
\bibliography{bibtex}

\end{document}